%% file: ms.tex
\documentclass[twocolumn]{aastex63}
\usepackage{subfigure}
\received{2021 February 28}
\revised{2021 March 8}
\accepted{2021 March 9}
\shorttitle{TESS observations of TX Col}
\shortauthors{Rawat et al.}
\hypersetup{
    colorlinks=true,
    linkcolor=blue,
    filecolor=magenta,      
    urlcolor=red,
    citecolor=blue,
}
\begin{document}
\title{ TESS observations of TX Col: Rapidly varying accretion flow}
\correspondingauthor{Nikita Rawat}
\email{nikita@aries.res.in}

\author{Nikita Rawat}
\affil{Aryabhatta Research Institute of Observational sciencES (ARIES), Nainital 263001, India}
\affil{Deen Dayal Upadhyaya Gorakhpur University, Gorakhpur 273009, India}

\author{J. C. Pandey}
\affil{Aryabhatta Research Institute of Observational sciencES (ARIES), Nainital 263001, India}

\author{Arti Joshi}
\affil{Aryabhatta Research Institute of Observational sciencES (ARIES), Nainital 263001, India}
\affil{School of Physics and Technology, Wuhan University, Wuhan 430072, China}

\begin{abstract}
Using the first long-term photometry from the Transiting Exoplanet Survey Satellite, we have carried out a detailed time-resolved timing analysis of an intermediate polar TX Col. The power spectra of almost 52 days continuous time-series data reveal the orbital period of $5.691\pm 0.006$ hr, spin period of $1909.5\pm 0.2$ s, and beat period of $2105.76\pm 0.25 $ s, which is consistent with the earlier results. We have also found the presence of quasi-periodic oscillations (QPOs) for a few days with a period of 5850-5950 s, which appears to be due to the beating of the Keplerian period with the spin period of the white dwarf. The continuous data allowed us to look thoroughly at the day-wise evolution of the system's accretion geometry. We report here that the TX Col changes its accretion mechanism even on a time-scale of one day, confirming its variable disk-overflow accretion nature. For the majority of the time, it was found to be disk-overflow system with stream-fed dominance, however, pure disk-fed and pure stream-fed accretions cannot be ruled out during the observations.
\end{abstract}

\keywords{Cataclysmic variable stars (203), Semi-detached binary stars(1443), DQ Herculis stars (407)}
 
\section{Introduction} \label{sec:intro}
Magnetic cataclysmic variables (MCVs) are semi-detached interacting binary systems, in which the primary white dwarf (WD) accretes matter from a Roche-lobe-filling secondary red dwarf star. Based on the magnetic field strength of the primary, MCVs are classified into two classes: polars and intermediate Polars (IPs). Polars are synchronous systems for which the magnetic field strength of the WD is more than 10 MG. The high magnetic field of WD prevents the formation of an accretion disk and the matter is accreted through magnetic field lines to the pole caps, while IPs are asynchronous systems and generally have magnetic field strength in the range of 1-10 MG. Hence, the material from the secondary star is accreted either through an accretion disk or an accretion stream, or a combination of both. For the majority of IPs, rotation period ($P_{\omega}$) is typically one-tenth of the orbital period ($P_{\Omega}$); i.e. $P_{\omega}\sim 0.1$ $P_{\Omega} $ \citep{1991MNRAS.248..370W}. 

In the IPs, three main accretion mechanisms are believed to occur, viz disk-fed (DF), stream-fed (SF), and disk-overflow (DO) accretion. In the DF accretion, the magnetic field strength of the WD truncates the inner edge of the accretion disk, forms ``accretion curtains," and dumps all the material to the magnetic poles of the WD \citep{1988MNRAS.231..549R}. In this case, the magnetosphere is smaller than the radius of the minimum approach of the freefalling accretion stream and the stream could then circle the white dwarf and form a disk, ignoring the feeble magnetic field. In the disk-less or SF accretion, the magnetic field is so strong that an accretion disk does not form, and material flows along the magnetic field lines to the pole caps \citep{1986MNRAS.218..695H}. Whereas in the case of a disk-overflow accretion, disk is present, but a part of the accretion stream skips the disk and directly interacts with the magnetosphere, leading to both disk-fed and stream-fed accretion simultaneously \citep{1989ApJ...340.1064L, 1996ApJ...470.1024A}. In another possibility of diamagnetic blob accretion, both disk-fed and stream-fed accretion can occur, but the difference here is that accretion flow is regarded as lumpy and diamagnetic \citep{1993MNRAS.261..144K, 1995MNRAS.275....9W}. In the disk-fed and disk-overflow models, the accretion material impacts both magnetic poles, while in the case of the only stream-fed, it impacts a single pole at any instant, but will flip between the two poles on the beat period between the spin and orbital periods of the system \citep{2001cvs..book.....H}. In all of the above models, optical modulation at different periods is seen either due to the reprocessing of the high-energy X-ray beams from the WD polar caps or due to the internal energy dissipation inside the disk. Absorption of X-rays by a structure locked to the WD, such as accretion curtain and accretion disk, results in an optical spin pulse, and reprocessing of X-ray beams by the structure locked to the binary orbit such as bright spot produces an optical beat pulsation \citep[see][] {1986MNRAS.219..347W}. The modulation at the orbital period is caused by obscurations of the WD by material rotating in the binary frame.
\par Based on all the above described models, the mode of accretion in IPs can be explained with the presence of the spin ($\omega$), beat ($\omega - \Omega$), and sideband frequencies in the power spectrum of a time-series data. The DF accretion gives rise to the modulation at the spin frequency of the white dwarf \citep{1995A&A...298..165K}. While in the case of SF accretion, modulations occur at the lower orbital sideband of the spin frequency \citep{1992MNRAS.255...83W}. For a disk-overflow accretion, modulations at both $\omega$ and $\omega-\Omega$ frequencies are expected to occur, and the main difference lies in the varying amplitude of modulation between the two. The modulation at $\omega-2\Omega$ originates as a sideband generated by orbital modulation of the $\omega-\Omega$ \citep[see][] {1986MNRAS.219..347W}, which confirms the presence of $\omega-\Omega$ frequency. Further, the orbital modulation of the spin frequency gives rise to $\omega+\Omega$ frequency. In both cases, this also alters the observed power at spin and beat frequencies, respectively.

\subsection{TX Col}
TX Col (=1H0542-407) was discovered by the HEAO-1 satellite and identified as an IP by \citet{1986ApJ...311..275T}. \citet{1989ApJ...344..376B} found the $P_{\Omega}$ and $P_{\omega}$ of 5.7 hr and 1914 s using the optical and X-ray observations. From extensive high-speed U, B, V, R, and I photometry, they also found a dominant period at 2106 s, which is the beat period of the system. Using ASCA and ROSAT observations, TX Col was found to be accreting predominantly via a disk in 1994, while in 1995 a substantial amount of the  accretion was occurred through the stream \citep[see][]{1997MNRAS.289..362N}. Further, the relative strength of the beat and spin modulation was found to be highly variable in the system in between the observations even on a timescale of a month \citep{1999ASPC..157...47W}. From the extensive photometry spanning $\sim$ 12 yr, \citet{2007MNRAS.380..133M} reported the detection of quasi-periodic oscillations (QPOs) of $\sim$ 5900s, for which they favored the mechanism where QPOs result from the beating of the Keplerian period of the orbiting blobs with the spin period of WD. Recently, Pandey et al. (2021, in preparation) also found its variable disk-overflow accretion nature using Chandra, Swift, and Suzaku observations. Hence, TX Col is an interesting IP to study further as it changes its accretion geometry frequently.

\par In this paper, we present a detailed investigation of the optical photometry of TX Col as observed from the Transiting Exoplanet Survey Satellite (TESS) mission, spanning a range of 52 days with a cadence of 2 minutes. The beauty lies in the capabilities of the continuous short-cadence observations by TESS. For the first time, the continuous observations allowed us to probe the system's accretion geometry in a very short cadence of one day.

The paper is organized as follows. In Section \ref{sec:obs}, we describe the observations and data, and analysis and results are described in Section \ref{sec:lc}. Finally, we present discussion in \ref{sec:disc} and summary and conclusions in section \ref{sec:conc}.

\section{Observations and data} \label{sec:obs}
TESS observed TX Col with the camera 3 during sectors 5 and 6 between 2018 November 15 and 2019 January 6. The TESS satellite has four telescopes with four cameras as backend instruments. All cameras with field of view of each 24$\times$24 deg$^2$  are aligned to cover $24\times90$ degree strips of the sky called ‘sectors' \citep[see][for details]{2015JATIS...1a4003R}. The TESS bandpass extends from 600 to 1100 nm and the sensitivity peaks near 900 nm and drops rapidly at wavelengths longer than 1000 nm. The observations were continuous with the exception of almost 4 days between the two sectors. The data were stored under the Mikulski Archive for Space Telescopes with identification number `TIC 21505340'. The data taken during an anomalous event had quality flags greater than 0 in the FITS file data structure, and thus we have considered only the data with the ``quality flag" = 0. For our analysis, we have taken ``PDCSAP" flux, which is the simple aperture photometry flux after removing the common instrumental systematics from it using the cotrending basis vectors. The PDCSAP flux also corrects for the amount of flux captured by the photometric aperture and crowding from known nearby stars\footnote{https://archive.stsci.edu/tess/}.

\begin{figure*}
\centering
\subfigure[]{\includegraphics[width=\textwidth]{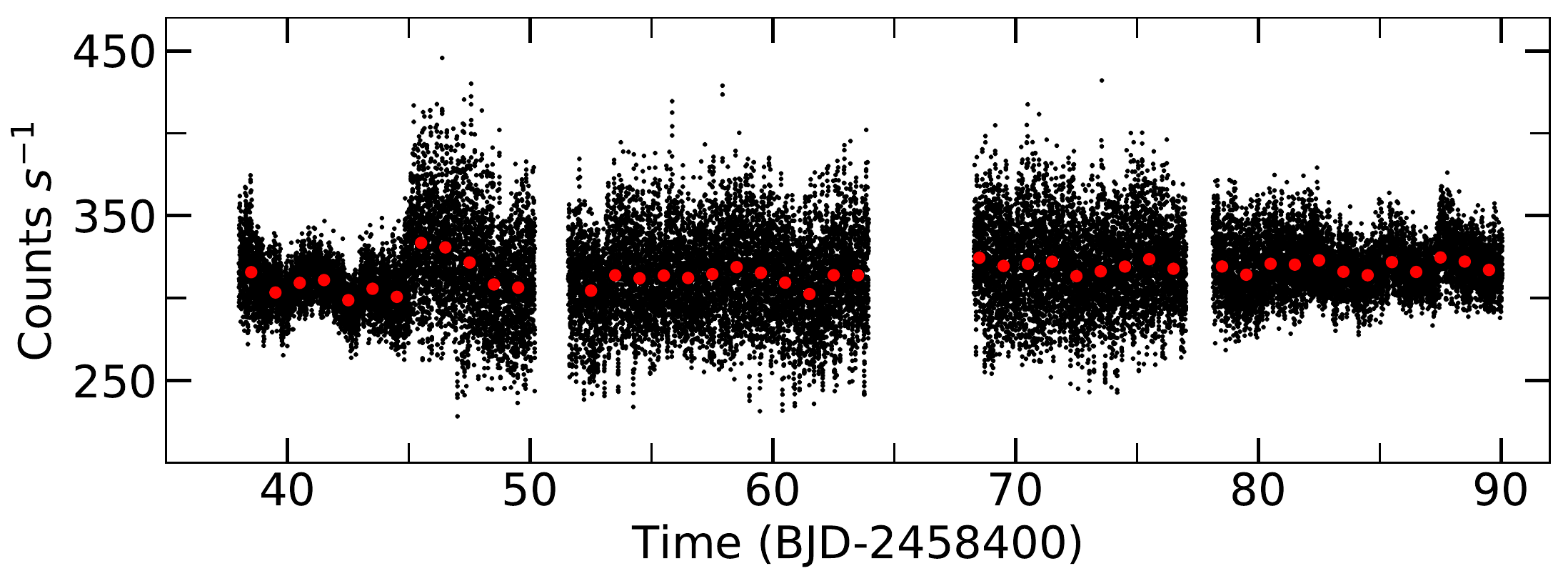}\label{fig:lc}}
\subfigure[]{\includegraphics[width=\textwidth]{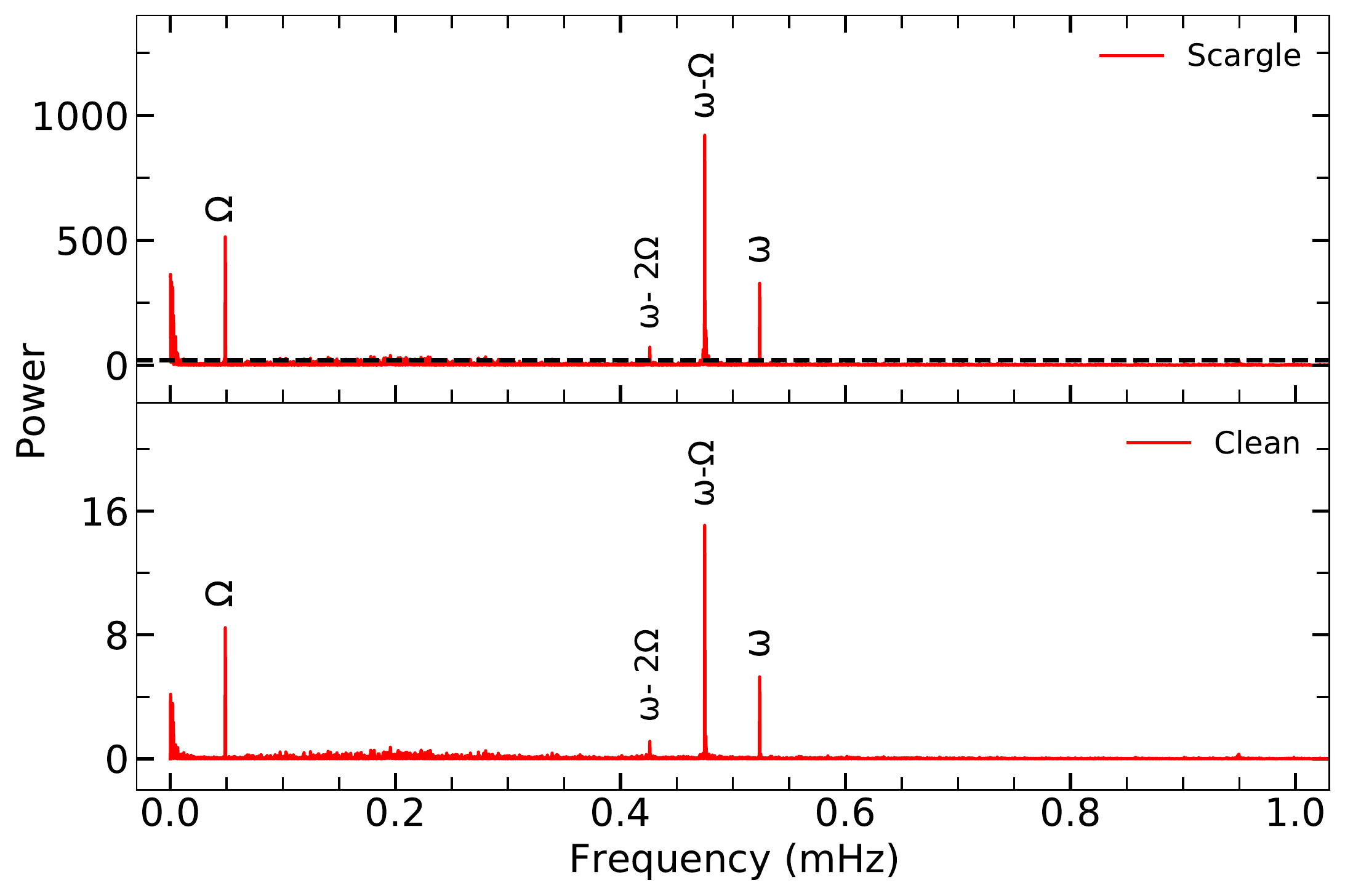}\label{fig:ls}}
\caption{(a) TESS light curve of TX Col, where the red dots represent the mean counts $s^{-1}$ of each day. (b) Lomb-Scargle power spectra (top panel) and CLEANed power spectra (bottom panel). The horizontal dashed line in the top panel represents the 90\% confidence level.}
\end{figure*}

\section{Analysis and results} \label{sec:lc}
\subsection{Light Curve}
Figure \ref{fig:lc} shows the TESS light curve of TX Col, and variability in the light curve is seen clearly. During the initial days of the observations from day 38 to 46, the variability amplitude is small with respect to the observations from day 46 to 77, whereas toward the end of the observations, amplitude of variability is found to have decreased again. We have performed the periodogram analysis using the Lomb-Scargle (LS) periodogram  method \citep{1976Ap&SS..39..447L,1982ApJ...263..835S}. The LS power spectrum of the complete dataset is shown in Figure \ref{fig:ls}. Several significant peaks are detected in the power spectrum. The significance of these detected peaks is determined by calculating the false-alarm probability (FAP; \citet{1986ApJ...302..757H}). The horizontal dashed line in Figure \ref{fig:ls} represents the 90\% confidence level. The dominant peaks are marked as $\Omega$, $\omega$, $\omega-\Omega$, and $\omega-2\Omega$. We have also employed the CLEAN algorithm given by \citet{1987AJ.....93..968R} to confirm the periodicity further. The CLEAN algorithm helps in removing the complex window function from the power spectrum that is produced due to the gaps in data segments. The bottom panel of Figure \ref{fig:ls} shows the CLEANed power spectrum of the combined TESS data. The CLEANed power spectrum was obtained with a loop gain of 0.1 and 1000 number of iterations. The number of iterations and loop gain were optimized to obtain a CLEANed power spectrum that is free of effects of spectral leakage. Similar to the LS periodogram, the peaks corresponding to the frequencies $\Omega$, $\omega$, $\omega-\Omega$, and $\omega-2\Omega$ are also found in the CLEANed power spectrum. Therefore, from the power spectral analysis of the complete dataset, we derived the orbital (P$_\Omega$), spin (P$_\omega$), and beat (P$_{\omega-\Omega}$) periods as $5.691\pm 0.006$ h, $1909.5\pm0.2$ s and $2105.76\pm0.25$ s, respectively.

\subsection{One-day time-resolved power spectral analysis}
\begin{figure*}
\gridline{\fig{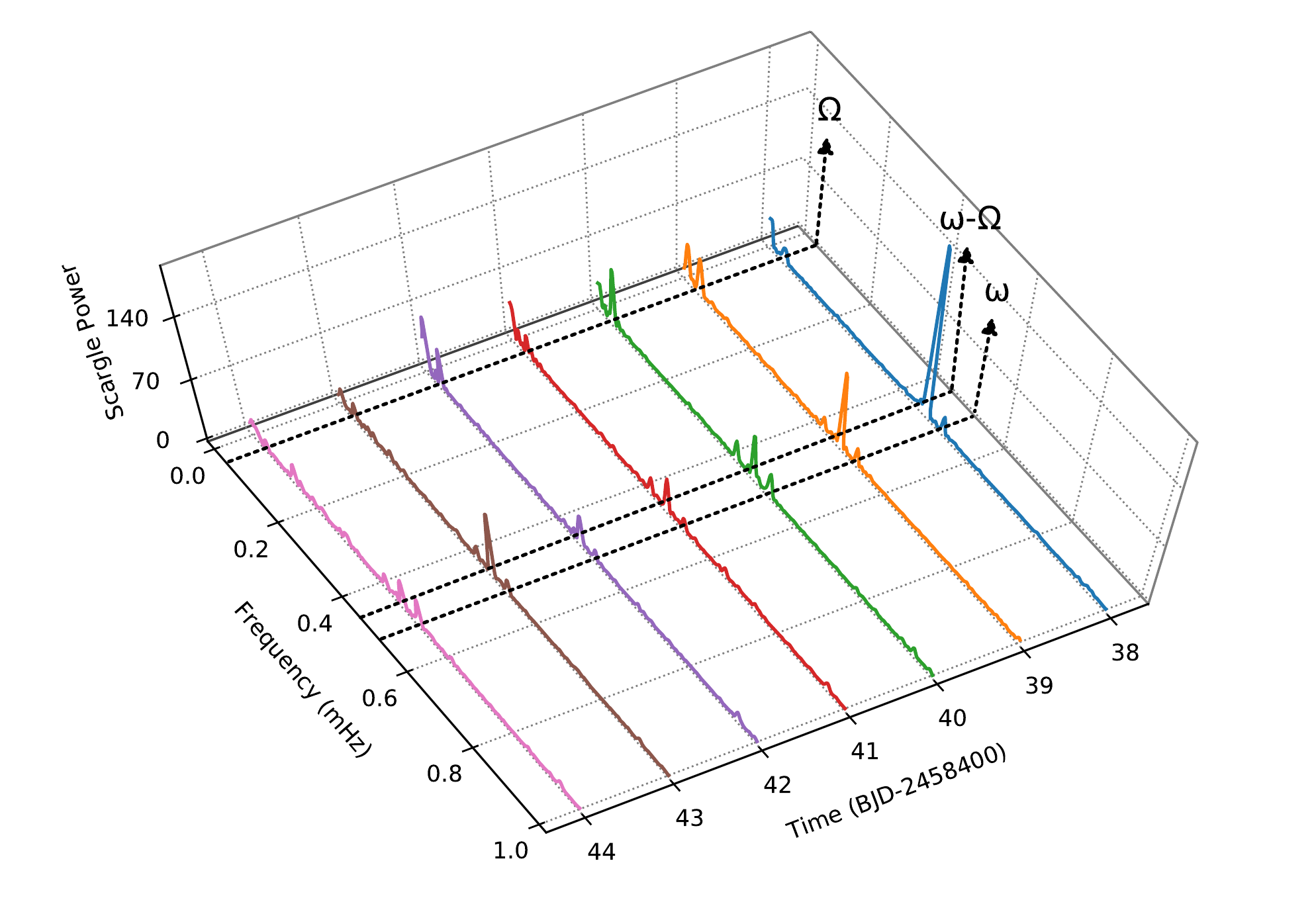}{0.5\textwidth}{}
          \fig{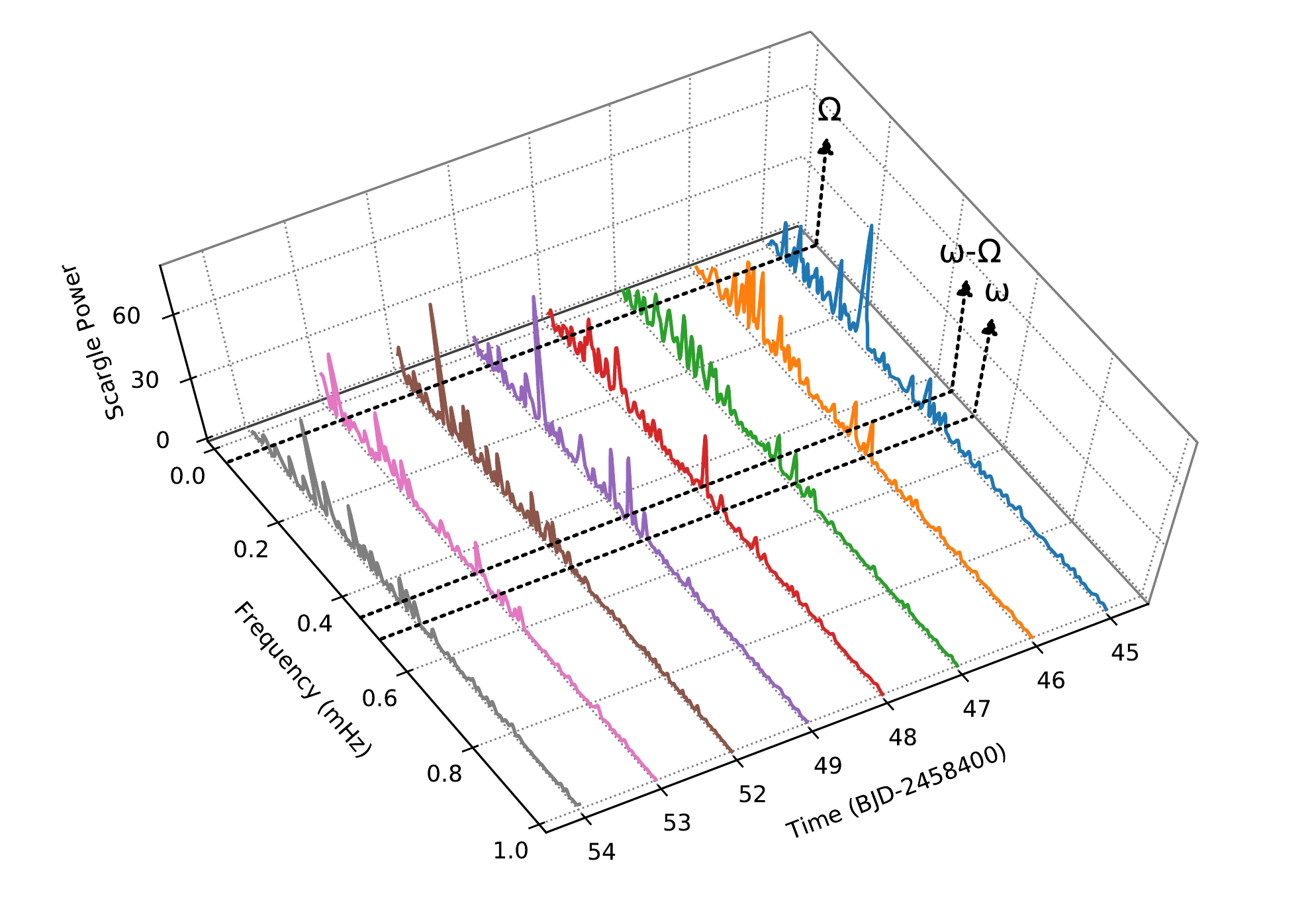}{0.5\textwidth}{}
           }
\gridline{ \fig{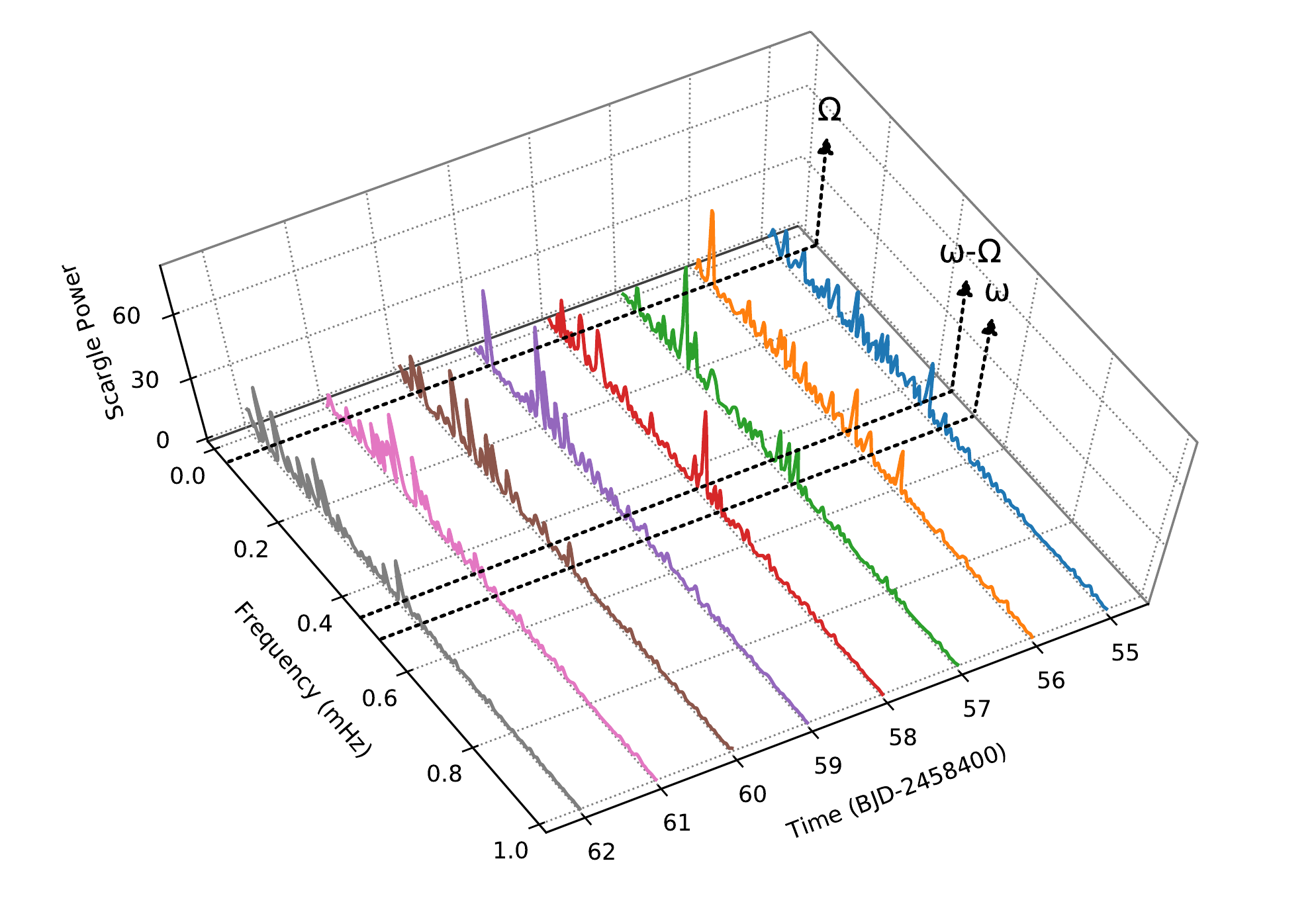}{0.5\textwidth}{}
           \fig{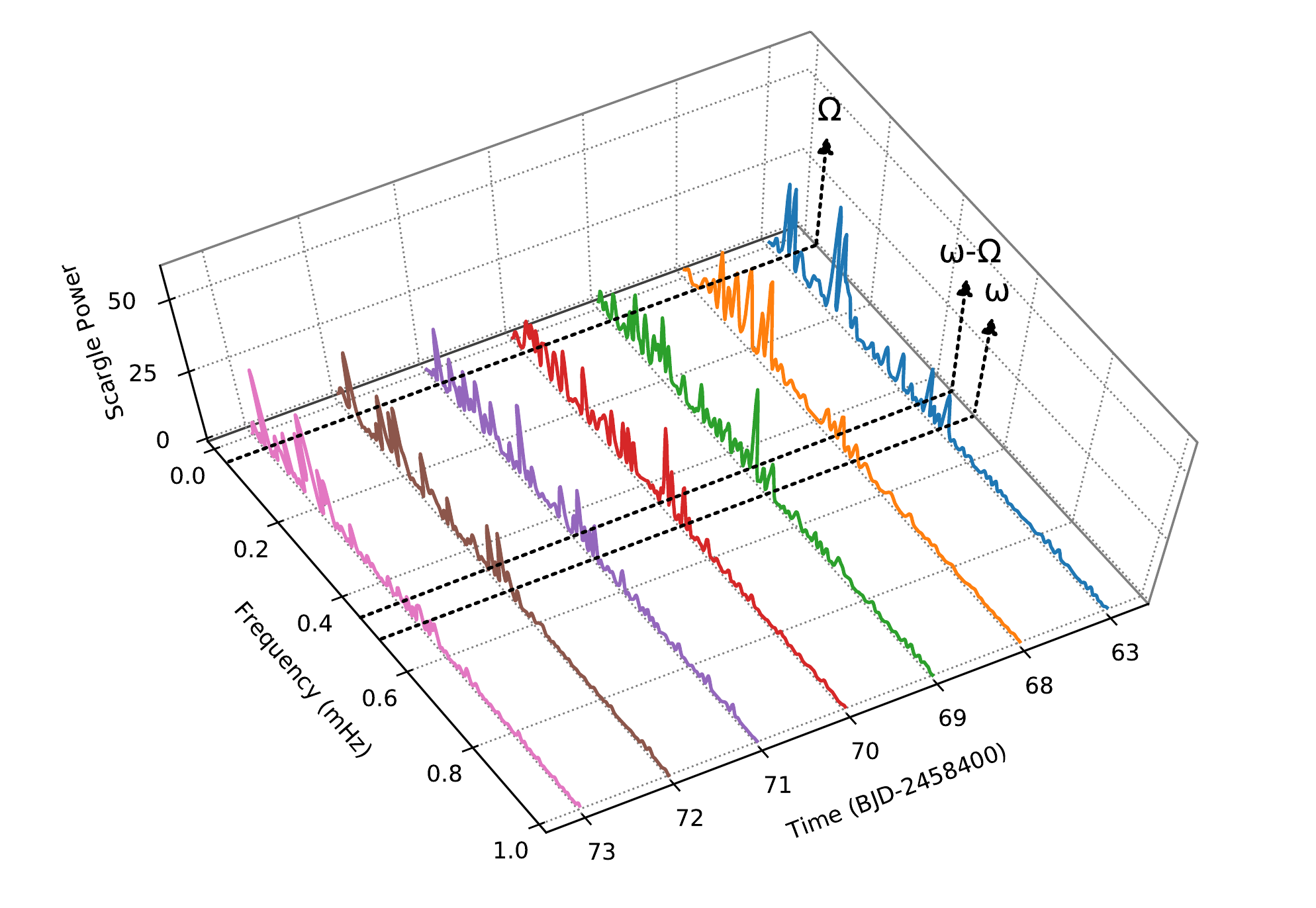}{0.5\textwidth}{}
            }
\gridline{ \fig{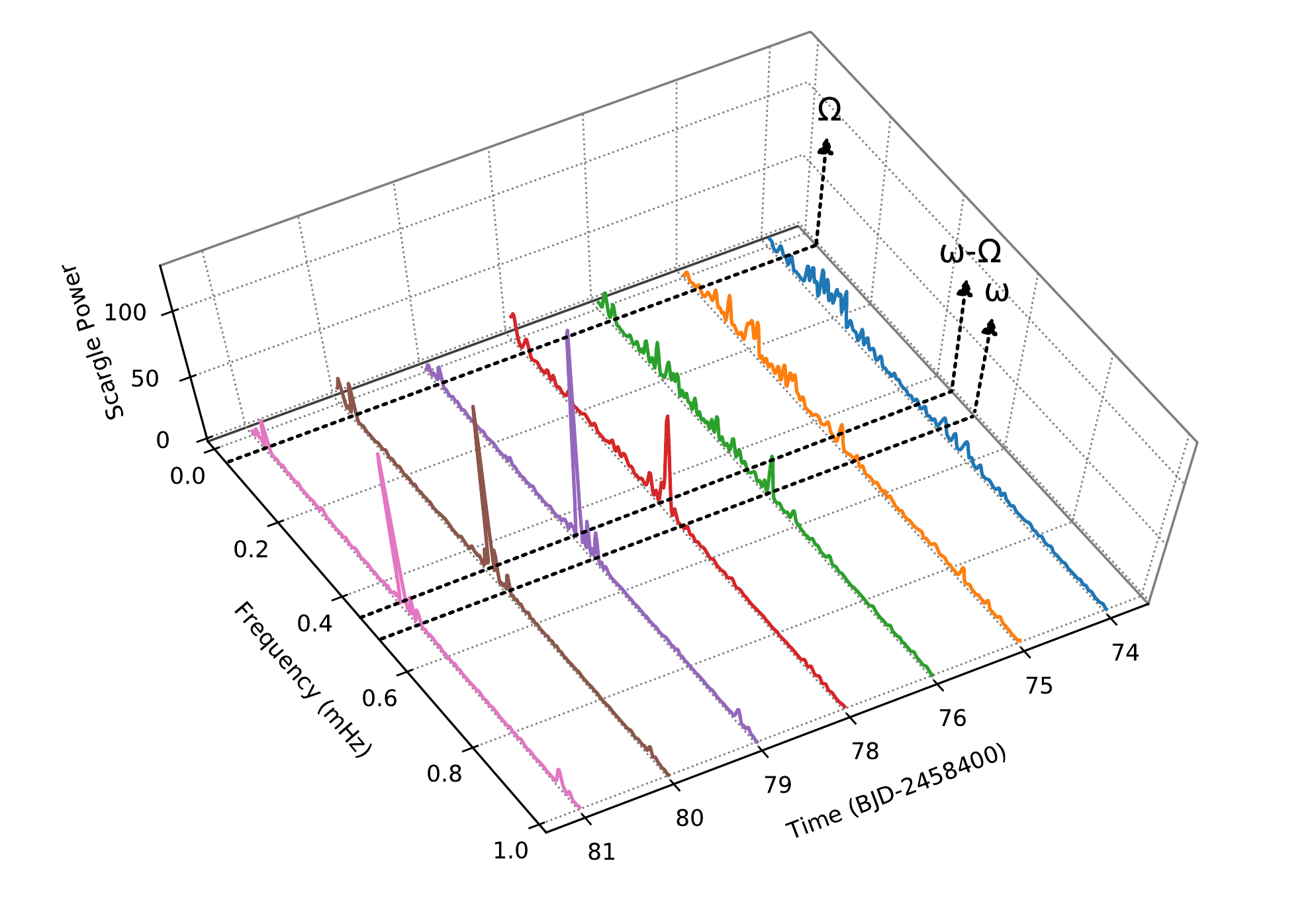}{0.5\textwidth}{}
           \fig{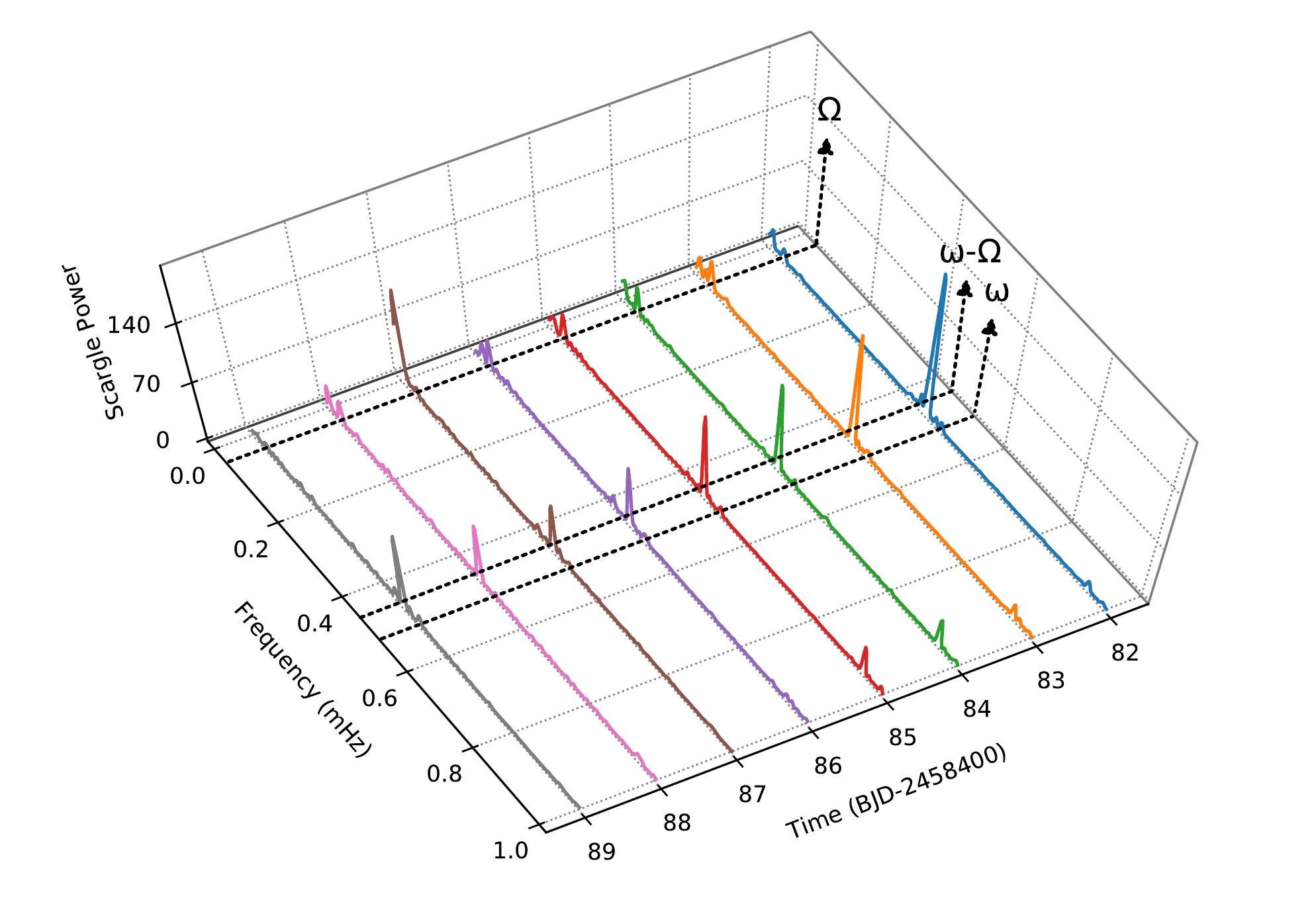}{0.5\textwidth}{}
            }
\caption{Power spectra of TX Col for each epoch of observations where the expected location of orbital, beat, and spin frequencies are also shown. } \label{fig:power_spec}
\end{figure*}

Intending to trace the change in the accretion geometry of TX Col, we have divided the complete dataset into one-day time segments, in which each data segment covers almost four orbital cycles of TX Col. These one-day time segment data were taken to better constrain the beat, spin, and orbital periods of the system. In this way, we could make 45 data segments for further timing analysis. Each data segment was subjected to an LS periodogram to search for the frequencies and variation in the power of each significant frequency. In Figure \ref{fig:power_spec}, we have shown the power spectra of each segment, where $\Omega$, $\omega$, and $\omega-\Omega$ frequencies are also labeled. Using the FAP, we have quoted the peaks with a 90\% confidence level in the power spectrum for each day. The significant frequencies are also given in Table \ref{tab:amp_mechanism}.

\par
The power spectrum of TX Col was found to be changing from one day to another day during the entire duration of TESS observations. If we examine Figure \ref{fig:power_spec}, we notice that both spin and beat frequencies were present with varying powers for most of the days. There were exceptions like for days 60 and 73, where the beat frequency was found to be completely absent, whereas for days 53, 54, and 78 no power at the spin frequency was observed. In the power spectrum, we have also noticed that for days 59, 74, and 76, and for days 45, 52, 55, 59, 61, 62, 68, 72, 75, 83, 84, 86, 87, and 88, beat and spin frequencies, respectively, were detected but they all lie above the 50\% and below the 90\% confidence level. For days 44, 46, 47, and 57, we observed that the beat and spin frequencies have similar powers. For days 39, 40, 41, 43, 45, 49, 52, 62, 71, 78, 86, and 87, a significant power at $\omega-2\Omega$ was also found. The second harmonic of $\omega-\Omega$ was detected with varying power for the days 42, 79, and 81 to 85, whereas the third harmonic of $\omega-\Omega$ frequency was observed on days 39, 83, and 86. Further, on day 73, we have found $\omega+\Omega$ frequency in the power spectrum just around at the 90\% confidence level. In our detailed one-day power spectral analysis, we have also found the frequencies corresponding to the periods of $\sim$ 5850-5950 s on days 46, 52, and 71. 

\subsection{Phased Light Curve}
\begin{figure*}[h]
\gridline{
          \fig{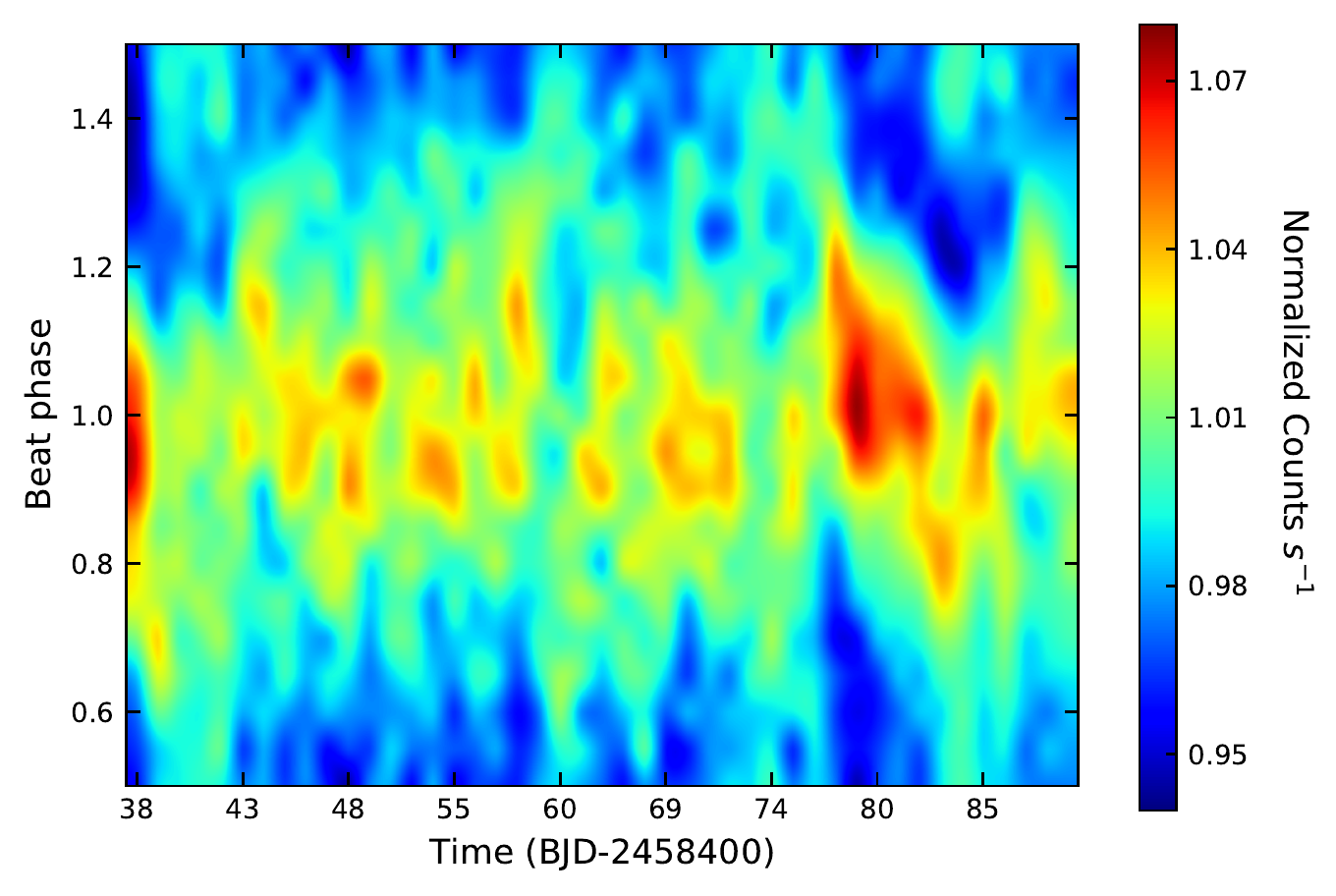}{0.5\textwidth}{(a)}
          \fig{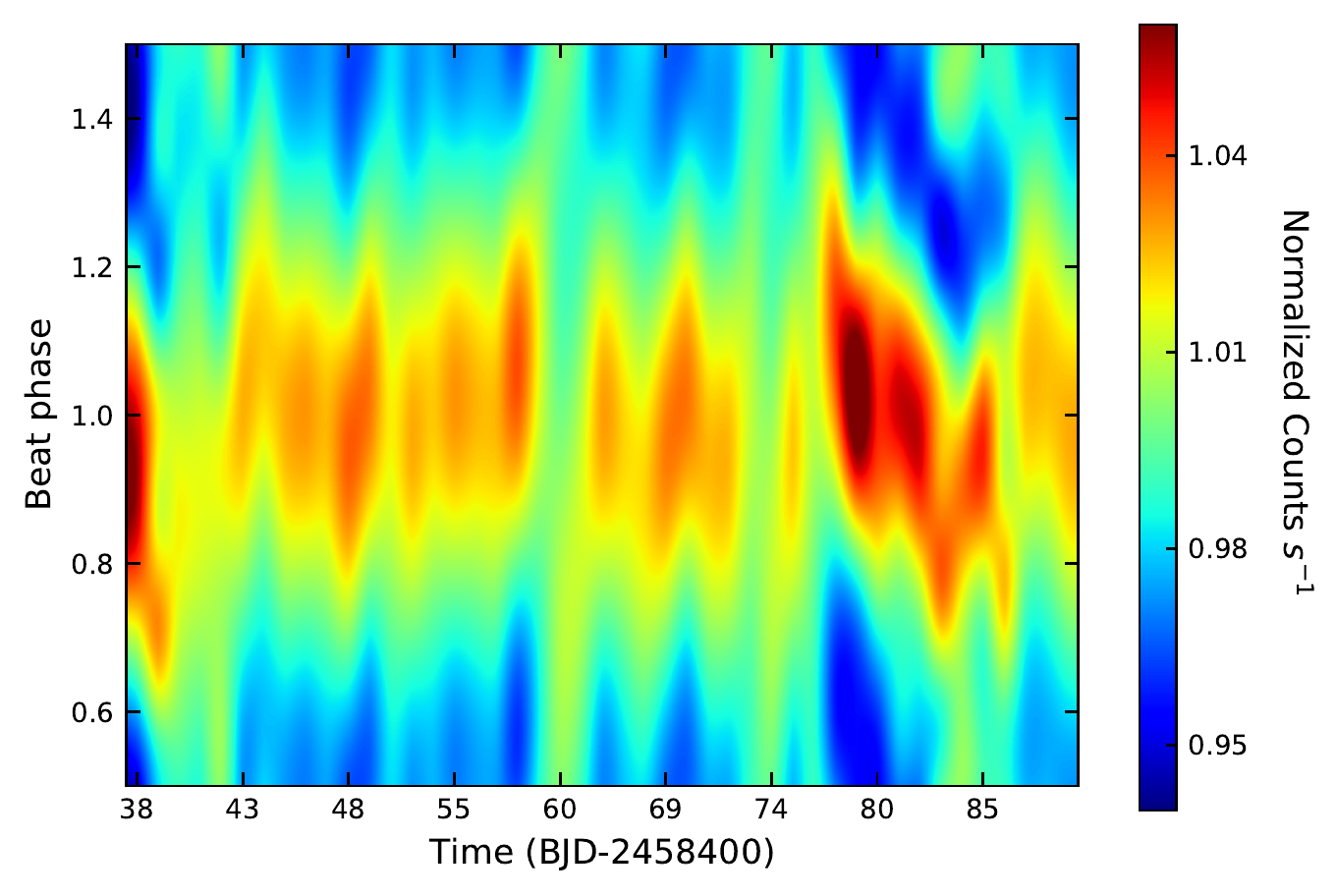}{0.5\textwidth}{(b)}
 }
\gridline{
          \fig{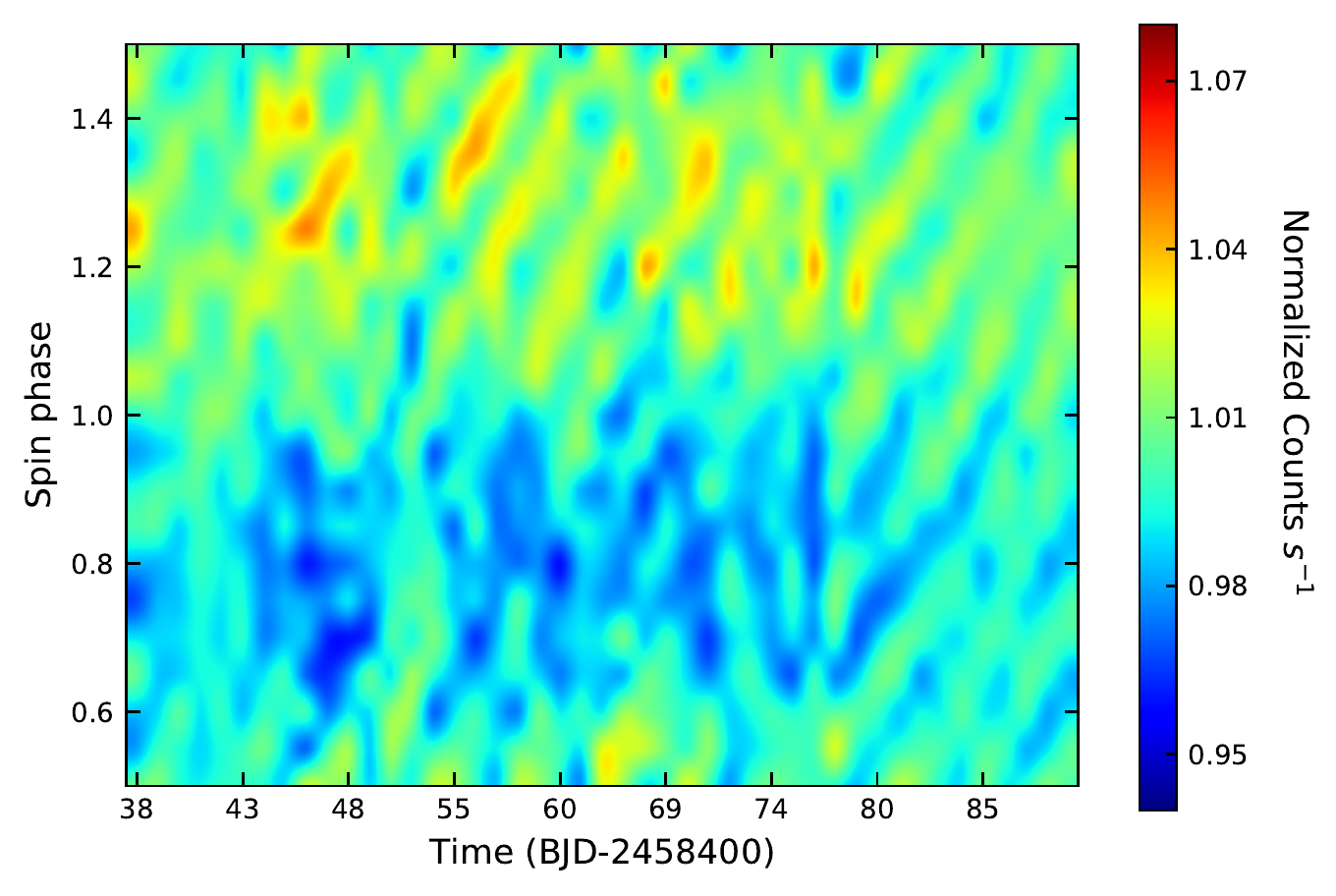}{0.5\textwidth}{(c)}
          \fig{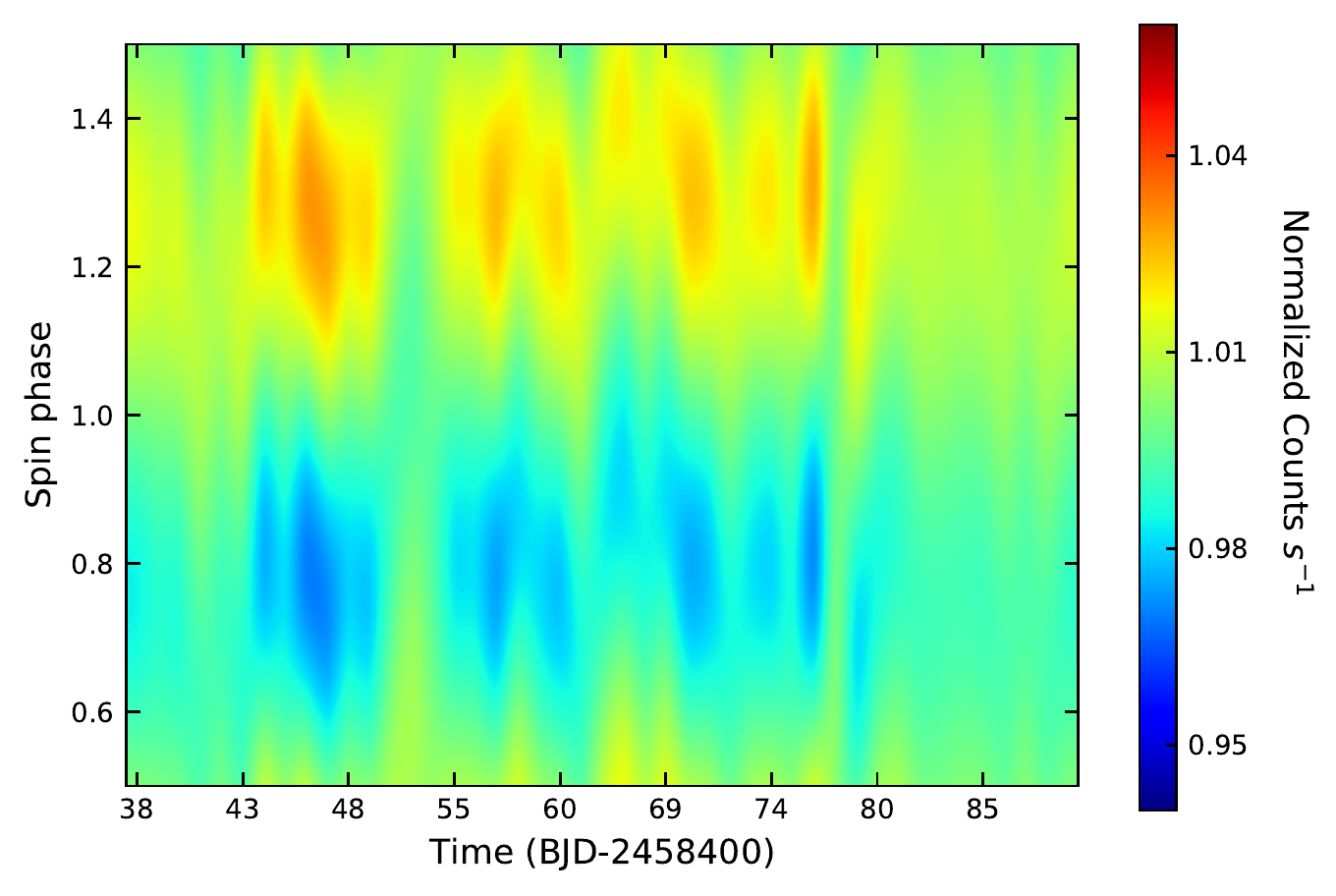}{0.5\textwidth}{(d)}
 }
\caption{Phased light curves of TX Col for the beat and spin periods\label{fig:amp_modu}. Left panels (a) and (c) are observed, whereas the right panels (b) and (d) are obtained after the best-fit sinusoidal curves.}
\end{figure*}

\begin{figure*}
    \centering
    \includegraphics[width=17cm,height=20cm]{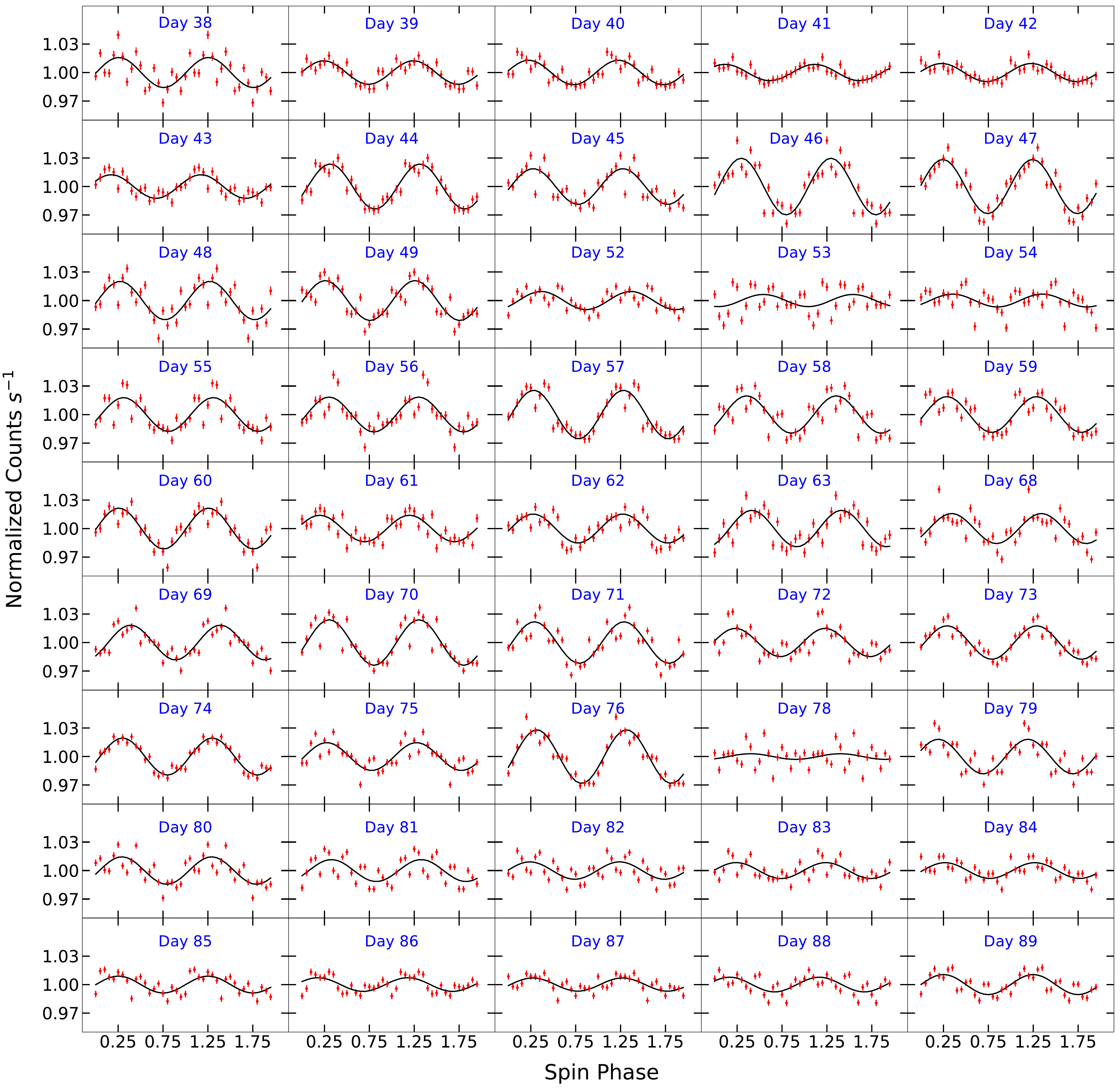}
    \caption{Spin phase-folded light curves for each day, where black solid curve represents the best-fit sinusoidal to the folded light curves.}
    \label{fig:spin_folded}
\end{figure*}

\begin{figure*}
    \centering
    \includegraphics[width=17cm,height=20cm]{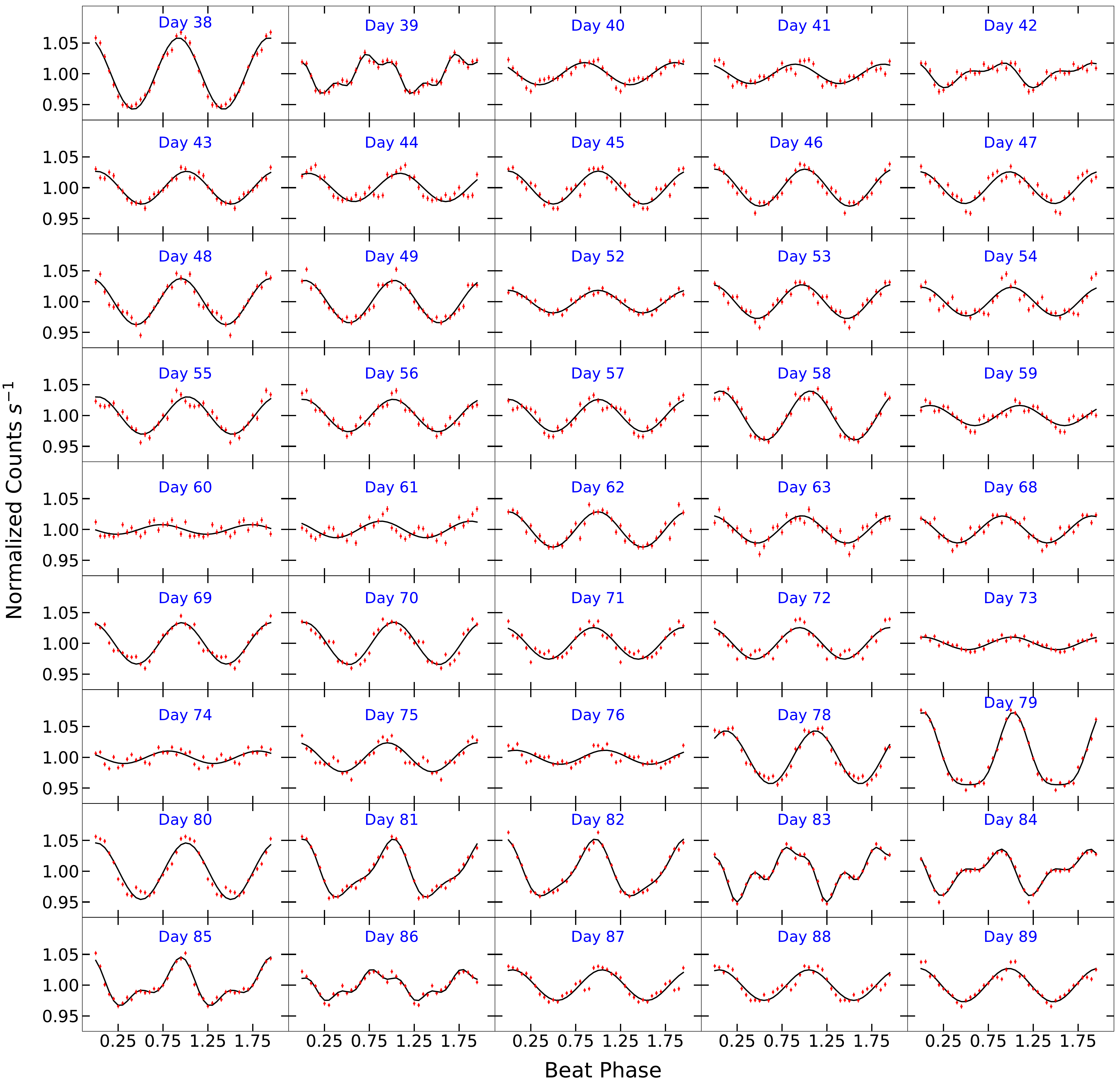}
    \caption{Beat phase-folded light curves for each day, where the black solid curve represents the best-fit sinusoidal to the folded light curves. For days 39, 42, 79, 81, 82, 83, 84, 85, and 86, the best fit was obtained by fitting the combination of two or three sine curves with frequency ${\omega-\Omega}$ and its harmonics. Also on these days, apart from ${\omega-\Omega}$ frequency, its second and/or third harmonics were present in the power spectrum. }
    \label{fig:beat_folded}
\end{figure*}

We have folded light curve of each day over spin (1909.5 s) and beat (2105.76 s) periods using the epoch HJD=2452290.286025, which was used to calculate spin ephemeris by \citet{2007MNRAS.380..133M}.  All the light curves were folded with the binning of 20 points in a phase. We have also derived the amplitude of modulations for the spin and beat pulsations with a best-fit sinusoidal function to the folded light curves of all one-day segments. Concisely, spin and beat phase-folded light curves are represented as the color composite plots (see the left panels of Figure \ref{fig:amp_modu}). However, the best-fitted sinusoidal data corresponding to spin and beat phases are plotted in the right set of panels of Figure \ref{fig:amp_modu}. Fig.\ref{fig:amp_modu}(a) and \ref{fig:amp_modu}(b) represent the dominance of the beat modulation over the spin modulation for the majority of the days, which was also confirmed by the power spectral analysis of each day (see Fig.\ref{fig:power_spec}). We have also plotted the spin and beat phase-folded light curves for all epochs of observations that are shown in Figures \ref{fig:spin_folded} and \ref{fig:beat_folded}, respectively. In some beat folded light curves, a single sinusoidal function did not fit well; therefore, an additional sinusoidal function with either a second or a third harmonic was fitted. Also, on these days either second or third harmonic of $\omega-\Omega$ frequency was present in the power spectra. We have obtained amplitude modulation for the spin and beat periods that are given in Table \ref{tab:amp_mechanism}. We have noticed that in the power spectra for days 59, 74, and 76, and for days 83, 84, 86, 87, and 88, the beat and spin pulses lie below the 90\% confidence level, whereas folded light curves showed a marginal modulation. Similarly, for days 45, 52, 55, 59, 61, 62, 68, 72, and 75, although the spin pulse was found to lie below the 90\% confidence level in the power spectra, a substantial amount of spin modulation has appeared on these days. Thus, on these days, we considered the presence of these modulations. Further, on days 60 and 73, the beat frequency was completely absent in the power spectra, which was also reflected in the beat folded light curves as a marginal modulation appeared to be present. Similarly, the power spectra of days 53, 54, and 78 have not shown the presence of spin frequency that was also confirmed by the spin folded light curves as the spin modulation did not seem to be present on these days.
The dominance of modulations has been decided by the ratio of spin ($A_{\omega}$) to beat ($A_{\omega-\Omega}$) amplitudes,  $A_{\omega}$/$A_{\omega-\Omega}$. If $A_{\omega}$/$A_{\omega-\Omega}$ $\sim$ 1.0, then both modulations were present equally. If $A_{\omega}/A_{\omega-\Omega}>$ 1.0, spin modulation was dominant, and if $A_{\omega}/A_{\omega-\Omega}<$ 1.0, then the beat modulation was dominant. Figure \ref{fig:ratio amp} shows the variation of $A_{\omega}$/$A_{\omega-\Omega}$ with time, and it is clear that for the majority of days beat amplitudes are higher than spin amplitudes.

\section{discussion} \label{sec:disc}
Using the power spectrum and amplitude modulation for each day, we now discuss the changes that have occurred in TX Col over the entire period of TESS observations. The accretion mechanism can be explained based on the combined effect of the presence of $\omega$ and $\omega-\Omega$ frequencies in the power spectrum and their amplitude of modulations along with the presence of the sideband frequencies. Based on these criteria, Table \ref{tab:amp_mechanism} shows the accretion mechanism for each day, where it is evident to see that the accretion mechanism sometimes has changed even in a duration of 1 day and sometimes it has not changed even in 3 to 4 days. For days 44, 46, 47, 57, 59, and 61, the ratio $A_{\omega}$/$A_{\omega-\Omega}$ was found to be $\sim$ 1.0, indicating the system is a disk-overflow with equal dominance of both stream- and disk-fed accretions. For the majority of days, the accretion mechanism was found to be disk-overflow with SF dominance (DOSF) but for two epochs, days 74 and 76, the system was found to be disk-overflow with DF dominance (DODF). If we investigate the beat folded light curves of the days 60 and 73, a marginal modulation seems to be present on these days. The presence of ${\omega+\Omega}$ on day 73, which is the upper orbital sideband of $\omega$ is additional confirmation for the presence of the spin pulse. Therefore, on two occasions, viz days 60 and 73 the accretion mechanism appears to be disk-fed. For the days 53, 54, and 78, $\omega$ frequency was found to be absent in the power spectra, which is also evident from the lack of spin modulation in the folded light curves. Therefore, it appears that on these days TX Col is accreting via stream-fed only. Also, the presence of ${\omega-2\Omega}$ on day 78, which is the lower orbital sideband of ${\omega-\Omega}$, further confirms the pure stream-fed accretion. The double-peaked-like structure is also seen in the beat folded light curves for days 39, 42, 79, and 81-86, which could be due to the presence of the second and/or third harmonic of ${\omega-\Omega}$ on these days.

\begin{figure*}
    \centering
    \includegraphics[width=10cm,height=6cm]{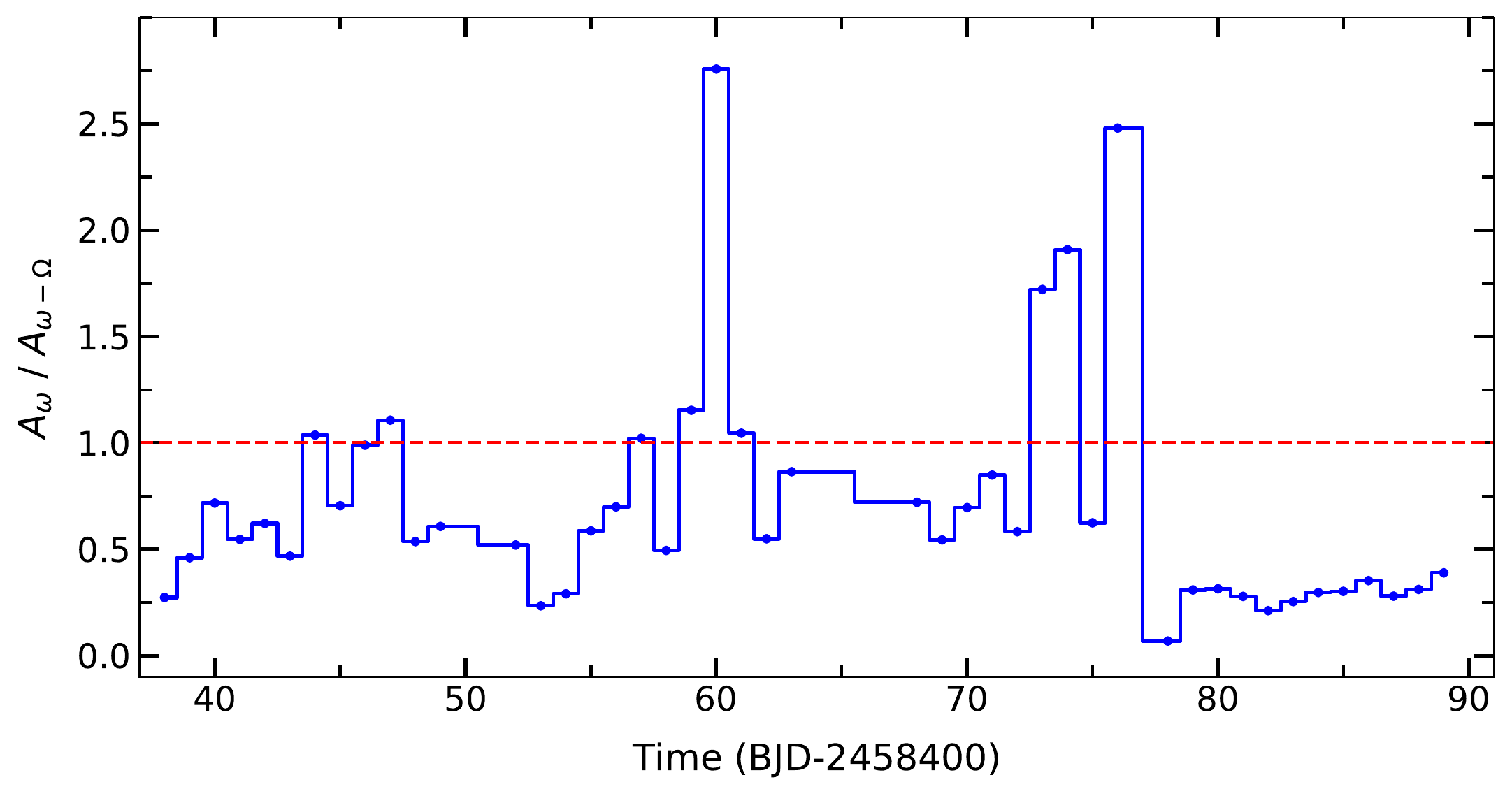}
    \caption{Ratio of spin to beat amplitudes obtained from the fitting for each day.}
    \label{fig:ratio amp}
\end{figure*}

If we inspect Figure \ref{fig:ratio amp}, we notice that the ratio $A_\omega/A_{\omega-\Omega}$ is found to be increasing from the beginning of the observations to the day 47, after that it is found to be decreasing until day 53. A similar trend is found from the day 53 to day 69 and from the day 69 to toward the end of observations. On days 60 and 76, the ratio $A_\omega/A_{\omega-\Omega}$ was found to have the highest values. Also, on the day 47, the ratio $A_\omega/A_{\omega-\Omega}$ is slightly above the value 1.0. From this, it appears that the spin modulation becomes dominant after every 13 - 16 days. Similarly, the minimum value of $A_\omega/A_{\omega-\Omega}$ was found to be on the days 38, 53, 69, and toward the end of the observations. It also appears that the ratio $A_\omega/A_{\omega-\Omega}$ becomes the minimum after every 15-16 days, which is similar to the maximum peak to peak values obtained. Therefore, from these observations, we infer that the system becomes a disk-dominant or stream-dominant accretor after every 13-16 days.

\subsection{Change in the mode of accretion}
Considering the different power spectra at an interval of one day, the system has changed its accretion mechanisms even on a time scale of a day during the observational period. This is the first time such a type of behavior has been noticed in any IP.
\par The system has shown all three types of accretion mechanisms varying from a disk-overflow to pure stream-fed or pure disk-fed on a time-scale of a few days, which is quite interesting because this directly tells us how rapidly the accretion rate is varying. Further, we have observed that the system is a variable disk-overflow with either stream-fed dominance or disk-fed dominance or an equal contribution from both stream-fed and disk-fed accretion. Among disk-overflow IPs, FO Aqr is a well-known source having accretion characteristics almost similar to TX Col. From past studies, the relative strength of the beat and spin signals were found to be varying on a time-scale of a few years, due to which pure disk-fed and disk-overflow accretion were observed in FO Aqr \citep{1996rftu.proc..123B}. Similar to FO Aqr \citep[see][]{1992MNRAS.254..705N, 1993MNRAS.265L..35H}, we have also found pure disk-fed accretion in some of the observations. However, in contrast to FO Aqr, for the first time, we have found the signature of pure stream-fed accretion in TX Col.\\
We will now revisit the possibilities given by \citet{1997MNRAS.289..362N} to account for the change in the accretion mechanism of TX Col:\\
1. They concluded that TX Col requires only a small change in mass flow to switch between the disk and disk-overflow geometries. We further add that the change in the mass flow is very outrageous, and despite the presence of the disk, sometimes the substantial amount of mass is accreting via stream only as ``stream-fed" accretion has been found in the 52 days of the observational period. It is well known that variable disk-overflow accretion takes place in TX Col \citep[see][]{1989ApJ...344..376B,1997MNRAS.289..362N, 1999ASPC..157...47W, 2002ASPC..261...92H}. Thus, a change in mass flow can be a reason for changing geometries between stream-fed dominance and disk-fed dominance. Further, since pure disk-fed accretion also takes place for some of the observational days, it implies that for those days, the contribution of mass accretion from the stream has been negligible in comparison to the disk.  \\
2. They suggested that change in accretion mode is related to changes in the disk itself, which can be easily understood from our one-day analysis of TX Col. The sudden changes between the disk-overflow to disk-fed or stream-fed cannot be only because of the change in the mass flow. The changes related to the disk, i.e. the change in the structure of the disk also takes place on such a time scale so that the combined effect of the two processes (i.e., the change in the mass accretion rate and the change in the structure of the disk) have shown such tremendous changes in the system.

\input{table_amp_mechanism}
\subsection{Presence of QPOs}
We have also found periods of $\sim$ 5850-5950 s for days 46, 52, and 71, and these periods could be the QPOs. The flow of matter in IPs can take the form of diamagnetic `blobs’ that orbit around the white dwarf \citep[see][for details]{1993MNRAS.261..144K, 1995MNRAS.275....9W}. The accretion flow in TX Col can be explained as a combination of a stream and orbiting blobs \citep{2002ASPC..261...92H}. Keeping these ideas in mind, \citet{2007MNRAS.380..133M} claimed that the QPO period results due to the beating of the Keplerian period of the orbiting `blobs’ with the spin period. The theoretical calculation of QPO period demands the calculation of Keplerian period ($P_{kep}=2 \pi r_{d}/v_{kep}$), which we have calculated in the following way. The mass of the WD in the TX Col system was determined in the range of 0.48-0.82 M\textsubscript{\(\odot\)} \citep[see Table 3 of][]{2010A&A...520A..25Y} in the different previous studies. Therefore, for our calculation, we have used the weighted mean of these mass values as $0.66 \pm 0.03$ M\textsubscript{\(\odot\)}, where the error on mass is the weighted standard deviation of different measurements. The WD radius, R\textsubscript{WD} was calculated to be $0.012\pm0.001$  R\textsubscript{\(\odot\)} using the mass-radius relation of \citet{1972ApJ...175..417N}. We have estimated  the mass of the secondary as $0.61 \pm 0.04$ M\textsubscript{\(\odot\)} using the mass–period relation given by \citet{1998MNRAS.301..767S}. As a result, the mass ratio (q) was estimated to be $0.92 \pm 0.07 $. The disk radius ($r_{d}$) lies in the range of 2-3 times $r_{r}$, where $r_{r}$ is the minimum outer radius of disk \citep{1980BAAS...12Q.749S}. The estimation of $r_{r}$ involves binary separation (a), which was calculated using the relations as given in \citet{1995CAS....28.....W}. The values of $r_{r}$ and `a' were found to be 1.08$\times10^{10}$ cm and 12.17$\times10^{10}$ cm respectively. In this way, $r_{d}$ was estimated to be in the range of (2.2-3.2)$\times10^{10}$ cm. Using these parameters of TX Col, we obtained a range for Keplerian velocity ($v_{kep}^2=G M_{WD}/r_{d}$) of 520-640 km $s^{-1}$. Further, we found that for the Keplerian velocity of $\sim 550$ km $s^{-1}$ and Keplerian radius at the outer disk of $\sim 2.5 \times 10^{10}$ cm, the QPO period comes out to be $\sim$ 5900 s which is similar to that obtained from periodogram analysis. The detection of the QPOs for some of the days confirms the change in the mass accretion rate as the orbiting ``blobs", which are released near the magnetosphere from where the material is accreted by the WD.

\section{Summary and Conclusions} \label{sec:conc}
The continuous and short-cadence observations of TX Col, obtained from TESS for a long period, allowed us to find changes occurring in the system on a time-scale of one day. Our main findings from this study are:
\begin{itemize}
\item[1.] The orbital, spin, and beat periods of TX Col are found to be consistent with the earlier studies.
\item[2.] TX Col is shown to accrete via disk-overflow accretion for $\sim$ 89\% of the total observing time and the other 11\% of the time either pure disk-fed or pure stream-fed. It exhibits disk-overflow with stream-fed dominance for 71.1\%, disk-overflow with equal dominance for 13.3\%, and disk-overflow with disk-fed dominance with only 4.4\% of the observing time. It also appears that TX Col has switched its accretion mechanism from disk-overflow to pure disk-fed accretion for two occasions during the observations. For the first time, a pure stream-fed accretion is also found to take place in this system for three different days.
\item[3.] We have found the QPOs of $\sim$ 5850-5950 s for three days, which appears to be due to the beating of the Keplerian period of the orbiting ``blobs" with the spin period as suggested by \citet{2007MNRAS.380..133M}. 
\item[4.] From these observations, it appears that the disk-dominant accretion and stream-dominant accretion repeat after every 13-16 days cycle. However, more observations are required to confirm this result. 
\end{itemize}
 
\noindent
\textbf{Acknowledgements:} We acknowledge the referee for reading our paper. This paper includes data collected with the TESS mission, obtained from the MAST data archive at the Space Telescope Science Institute (STScI). Funding for the TESS mission is provided by the NASA Explorer Program. STScI is operated by the Association of Universities for Research in Astronomy, Inc., under NASA contract NAS 5–26555.

\end{document}

%% file: table_amp_mechanism.tex
\startlongtable
\begin{deluxetable*}{cclllc}
	\tablecaption{Amplitudes of the Spin and Beat Modulations as Obtained from the Sine Fitting of the Phased Light Curves.    \label{tab:amp_mechanism}}
	\tablehead{\colhead{Day} & \colhead{Mean flux  }& \multicolumn{2}{c}{Amplitude (\%)} & \colhead{Frequencies} & \colhead{Accretion mechanism} \\
		\colhead{(BJD-2458400+)} & \colhead{counts s$^{-1}$}  & \colhead{$A_{\omega}$} & \colhead{$A_{\omega-\Omega}$}  & \colhead{present}
	}
	\startdata
	38 & 315.75 $\pm$ 0.31 & 1.6 $\pm$ 0.3 & 5.8 $\pm$ 0.2 & $\omega$, $\omega-\Omega$, $\Omega$ & DOSF \\
	39 & 303.42 $\pm$ 0.30 & 1.2 $\pm$ 0.1 & 2.7 $\pm$ 0.1 & $\omega$, $\omega-\Omega$,  $\Omega$, $\omega-2\Omega$, $3(\omega-\Omega)$ & DOSF \\
	40 & 309.25 $\pm$ 0.30 & 1.3 $\pm$ 0.2 & 1.8 $\pm$ 0.2 & $\omega$, $\omega-\Omega$,  $\Omega$, $\omega-2\Omega$ & DOSF \\
	41 & 310.95 $\pm$ 0.30 & 0.9 $\pm$ 0.1 & 1.6 $\pm$ 0.2 & $\omega$, $\omega-\Omega$,  $\Omega$, $\omega-2\Omega$ & DOSF \\
	42 & 298.76 $\pm$ 0.30 & 1.0 $\pm$ 0.1 & 1.5 $\pm$ 0.2 & $\omega$, $\omega-\Omega$,  $\Omega$, $2(\omega-\Omega)$ & DOSF \\
	43 & 305.69 $\pm$ 0.30 & 1.2 $\pm$ 0.1 & 2.7 $\pm$ 0.1 & $\omega$, $\omega-\Omega$,  $\Omega$, $\omega-2\Omega$ & DOSF \\
	44 & 300.79 $\pm$ 0.30 & 2.4 $\pm$ 0.2 & 2.3 $\pm$ 0.2 & $\omega$, $\omega-\Omega$,  $\Omega$ & DO\\
	45 & 333.48 $\pm$ 0.31 & 1.9 $\pm$ 0.2* & 2.7 $\pm$ 0.2 & $\omega-\Omega$,  $\Omega$, $\omega-2\Omega$ & DOSF \\
	46 & 330.78 $\pm$ 0.31 & 3.0 $\pm$ 0.3 & 3.0 $\pm$ 0.2 & $\omega$, $\omega-\Omega$ & DO \\
	47 & 321.60 $\pm$ 0.31 & 2.8 $\pm$ 0.2 & 2.6 $\pm$ 0.2 & $\omega$, $\omega-\Omega$,  $\Omega$ & DO \\
	48 & 308.31 $\pm$ 0.31 & 2.0 $\pm$ 0.3 & 3.8 $\pm$ 0.2 & $\omega$, $\omega-\Omega$ & DOSF \\
	49 & 306.34 $\pm$ 0.29 & 2.1 $\pm$ 0.2 & 3.4 $\pm$ 0.2 & $\omega$, $\omega-\Omega$,  $\Omega$, $\omega-2\Omega$ & DOSF \\
	52 & 304.54 $\pm$ 0.25 & 1.0 $\pm$ 0.2* & 1.8 $\pm$ 0.1 & $\omega-\Omega$,  $\Omega$, $\omega-2\Omega$ & DOSF \\
	53 & 313.88 $\pm$ 0.31 & 0.6 $\pm$ 0.3 & 2.7 $\pm$ 0.2 & $\omega-\Omega$, $\Omega$ & SF \\
	54 & 312.02 $\pm$ 0.31 & 0.7 $\pm$ 0.3 & 2.3 $\pm$ 0.3 & $\omega-\Omega$ & SF \\
	55 & 313.70 $\pm$ 0.31 & 1.8 $\pm$ 0.2* & 3.0 $\pm$ 0.2 & $\omega-\Omega$,  $\Omega$ & DOSF \\
	56 & 312.23 $\pm$ 0.30 & 1.8 $\pm$ 0.2 & 2.6 $\pm$ 0.2 & $\omega$, $\omega-\Omega$,  $\Omega$ & DOSF \\
	57 & 314.58 $\pm$ 0.31 & 2.6 $\pm$ 0.2 & 2.6 $\pm$ 0.2 & $\omega$, $\omega-\Omega$,  $\Omega$ & DO \\
	58 & 318.83 $\pm$ 0.31 & 2.0 $\pm$ 0.2 & 4.0 $\pm$ 0.2 & $\omega$, $\omega-\Omega$,  $\Omega$ & DOSF\\
	59 & 315.34 $\pm$ 0.31 & 1.9 $\pm$ 0.2* & 1.6 $\pm$ 0.2* &  $\Omega$ & DO \\
	60 & 309.42 $\pm$ 0.31 & 2.2 $\pm$ 0.2 & 0.8 $\pm$ 0.2 & $\omega$, $\Omega$ & DF \\
	61 & 302.47 $\pm$ 0.30 & 1.4 $\pm$ 0.2* & 1.3 $\pm$ 0.2 & $\omega-\Omega$ & DO \\
	62 & 313.93 $\pm$ 0.30 & 1.6 $\pm$ 0.3* & 2.9 $\pm$ 0.2 & $\omega-\Omega$,  $\Omega$, $\omega-2\Omega$ & DOSF \\
	63 & 313.86 $\pm$ 0.34 & 1.9 $\pm$ 0.2 & 2.2 $\pm$ 0.2 & $\omega$, $\omega-\Omega$,  $\Omega$ & DOSF \\
	68 & 324.47 $\pm$ 0.30 & 1.6 $\pm$ 0.3* & 2.2 $\pm$ 0.2 & $\omega-\Omega$ & DOSF \\
	69 & 319.54 $\pm$ 0.25 & 1.8 $\pm$ 0.2 & 3.4 $\pm$ 0.2 & $\omega$, $\omega-\Omega$,  $\Omega$ & DOSF \\
	70 & 320.79 $\pm$ 0.25 & 2.4 $\pm$ 0.2 & 3.4 $\pm$ 0.2 & $\omega$, $\omega-\Omega$,  $\Omega$ & DOSF \\
	71 & 322.13 $\pm$ 0.26 & 2.2 $\pm$ 0.2 & 2.6 $\pm$ 0.2 & $\omega$, $\omega-\Omega$,  $\Omega$, $\omega-2\Omega$ & DOSF \\
	72 & 313.33 $\pm$ 0.25 & 1.5 $\pm$ 0.2* & 2.6 $\pm$ 0.2 & $\omega-\Omega$,  $\Omega$ & DOSF \\
	73 & 316.27 $\pm$ 0.25 & 1.8 $\pm$ 0.1 & 1.0 $\pm$ 0.1 & $\omega$, $\Omega$, $\omega+\Omega$ & DF \\
	74 & 319.16 $\pm$ 0.25 & 1.9 $\pm$ 0.1 & 1.0 $\pm$ 0.2* & $\omega$ & DODF\\
	75 & 323.60 $\pm$ 0.25 & 1.5 $\pm$ 0.2* & 2.4 $\pm$ 0.2 & $\omega-\Omega$ & DOSF \\
	76 & 317.82 $\pm$ 0.25 & 2.8 $\pm$ 0.2 & 1.1 $\pm$ 0.2* & $\omega$,  $\Omega$ & DODF \\
	78 & 319.20 $\pm$ 0.28 & 0.3 $\pm$ 0.2 & 4.3 $\pm$ 0.2 & $\omega-\Omega$,  $\Omega$, $\omega-2\Omega$ & SF \\
	79 & 314.20 $\pm$ 0.25 & 1.8 $\pm$ 0.2 & 5.9 $\pm$ 0.1 & $\omega$, $\omega-\Omega$,  $\Omega$, $2(\omega-\Omega)$ & DOSF \\
	80 & 320.83 $\pm$ 0.25 & 1.4 $\pm$ 0.2 & 4.6 $\pm$ 0.2 & $\omega$, $\omega-\Omega$,  $\Omega$ & DOSF \\
	81 & 320.33 $\pm$ 0.25 & 1.2 $\pm$ 0.2 & 4.2 $\pm$ 0.1 & $\omega$, $\omega-\Omega$, $2(\omega-\Omega)$ & DOSF \\
	82 & 322.92 $\pm$ 0.25 & 0.9 $\pm$ 0.2 & 4.3 $\pm$ 0.1 & $\omega$, $\omega-\Omega$,  $\Omega$, $2(\omega-\Omega)$ & DOSF \\
	83 & 315.99 $\pm$ 0.25 & 0.8 $\pm$ 0.2* & 3.3 $\pm$ 0.1 & $\omega-\Omega$,  $\Omega$, $2(\omega-\Omega)$, $3(\omega-\Omega)$  & DOSF \\
	84 & 313.90 $\pm$ 0.25 & 0.8 $\pm$ 0.2* & 2.8 $\pm$ 0.1 & $\omega-\Omega$,  $\Omega$, $2(\omega-\Omega)$ & DOSF \\
	85 & 321.84 $\pm$ 0.25 & 0.9 $\pm$ 0.2 & 3.0 $\pm$ 0.1 & $\omega$, $\omega-\Omega$,  $\Omega$, $2(\omega-\Omega)$ & DOSF \\
	86 & 315.88 $\pm$ 0.25 & 0.7 $\pm$ 0.1* & 2.0 $\pm$ 0.1 & $\omega-\Omega$,  $\Omega$, $\omega-2\Omega$, $3(\omega-\Omega)$ & DOSF\\
	87 & 324.67 $\pm$ 0.25 & 0.7 $\pm$ 0.1* & 2.5 $\pm$ 0.2 & $\omega-\Omega$,  $\Omega$, $\omega-2\Omega$ & DOSF \\
	88 & 322.19 $\pm$ 0.25 & 0.8 $\pm$ 0.2* & 2.5 $\pm$ 0.2 &  $\omega-\Omega$,  $\Omega$ & DOSF \\
	89 & 317.10 $\pm$ 0.24 & 1.1 $\pm$ 0.2 & 2.7 $\pm$ 0.1 & $\omega$, $\omega-\Omega$ & DOSF \\
	\enddata
	\tablenotetext{}{The Accretion mechanism is decided on the basis of a combination of presence of $\omega$, $\omega-\Omega$, and sideband frequencies in the power spectrum on that particular day and the amplitudes of spin and beat modulations obtained from the fitting. DF: disk-fed, SF: Stream-fed, DOSF: disk overflow stream-fed dominance, DODF: disk overflow disk-fed dominance, DO: disk overflow with equal dominance}
	\tablenotetext{}{* represents the corresponding frequencies were below 90\% confidence level in the power spectrum, however the modulation was found in the folded light curves.}
\end{deluxetable*}